\documentclass[sigconf,nonacm]{acmart}

\AtBeginDocument{%
  }

\setcopyright{acmlicensed}
\copyrightyear{2018}
\acmYear{2018}
\acmDOI{XXXXXXX.XXXXXXX}

\acmConference[Conference acronym 'XX]{Make sure to enter the correct
  conference title from your rights confirmation email}{June 03--05,
  2018}{Woodstock, NY}

\acmISBN{978-1-4503-XXXX-X/2018/06}

\usepackage{booktabs}
\usepackage{multirow}
\usepackage{enumitem}
\usepackage{xcolor} 
\usepackage{booktabs}
\usepackage{multirow}
\usepackage{wrapfig}
\usepackage{graphicx}  
\usepackage{xcolor}    
\usepackage{listings}  
\usepackage{float}     
\usepackage{booktabs}  
\usepackage{subcaption}
\usepackage{graphicx}
\usepackage{xcolor}
\definecolor{diffgreen}{rgb}{0.0, 0.5, 0.0} 
\definecolor{diffred}{rgb}{0.6, 0.0, 0.0}   
\definecolor{codeblue}{rgb}{0.0, 0.0, 0.8}  
\usepackage[table]{xcolor}  
\usepackage{arydshln}       
\definecolor{codegreen}{rgb}{0,0.6,0}
\definecolor{codegray}{rgb}{0.5,0.5,0.5}
\definecolor{codepurple}{rgb}{0.58,0,0.82}
\definecolor{backcolour}{rgb}{0.95,0.95,0.92}

\lstdefinelanguage{JavaScript}{
  keywords={typeof, new, true, false, catch, function, return, null, catch, switch, var, if, in, while, do, else, case, break},
  keywordstyle=\color{blue}\bfseries,
  ndkeywords={class, export, boolean, throw, implements, import, this},
  ndkeywordstyle=\color{codegreen}\bfseries,
  identifierstyle=\color{black},
  sensitive=false,
  comment=[l]{//},
  morecomment=[s]{/*}{*/},
  commentstyle=\color{codegreen}\ttfamily,
  stringstyle=\color{red}\ttfamily,
  morestring=[b]',
  morestring=[b]"
}
\lstdefinestyle{mystyle}{
    backgroundcolor=\color{backcolour},
    commentstyle=\color{codegreen},
    keywordstyle=\color{magenta},
    numberstyle=\tiny\color{codegray},
    stringstyle=\color{codepurple},
    basicstyle=\ttfamily\footnotesize, 
    breakatwhitespace=false,
    breaklines=true,
    captionpos=b,
    keepspaces=true,
    numbers=left,
    numbersep=5pt,
    showspaces=false,
    showstringspaces=false,
    showtabs=false,
    tabsize=2,
    frame=single                     
}
\usepackage[framemethod=tikz]{mdframed}
\newenvironment{custommdframed}
  {\begin{mdframed}[style=customstyle]}
  {\end{mdframed}}
 \mdfdefinestyle{customstyle}{
  linecolor=gray!100,
  linewidth=3pt,
  innerleftmargin=3pt,
  topline=false,
  rightline=false,
  bottomline=false,
  leftline=true,
  innerrightmargin=3pt,
  innertopmargin=3pt,
  innerbottommargin=3pt,
  backgroundcolor=gray!15
}

\newcommand{\ourmodel}{\textsc{MemRepair}}

\begin{document}

\title{\ourmodel{}: Hierarchical Memory for Agentic Repository-Level Vulnerability Repair}

\author{Simiao Liu}
\affiliation{%
  \institution{Beihang University}
  \city{Beijing}
  \country{China}}
\email{buaalsm@buaa.edu.cn}

\author{Li Zhang}
\affiliation{%
  \institution{Beihang University}
  \city{Beijing}
  \country{China}}
\email{lily@buaa.edu.cn}

\author{Fang Liu}
\authornote{Corresponding author.}
\affiliation{%
  \institution{Beihang University}
  \city{Beijing}
  \country{China}}
\email{fangliu@buaa.edu.cn}

\author{Xiaoli Lian}
\affiliation{%
  \institution{Beihang University}
  \city{Beijing}
  \country{China}}
\email{lianxiaoli@buaa.edu.cn}

\author{Yang Liu}
\affiliation{%
  \institution{Beihang University}
  \city{Beijing}
  \country{China}}
\email{liuyang26@buaa.edu.cn}

\author{Yinghao Zhu}
\affiliation{%
  \institution{The University of Hong Kong}
  \city{Hong Kong}
  \country{China}}
\email{yhzhu99@gmail.com}

\renewcommand{\shortauthors}{Liu et al.}

\begin{abstract}

Modern software ecosystems face a rapidly growing number of disclosed vulnerabilities, increasing the need for automated repair techniques that can operate reliably at repository scale. Although Large Language Model (LLM)-based agents have recently shown promise for automated vulnerability repair (AVR), most existing systems still treat repair as a single generation step over the currently visible code context. As a result, they lack a persistent mechanism for reusing prior fixes or learning from failed validation attempts, which limits their effectiveness on complex, multi-file repair tasks. We present \ourmodel{}, a memory-augmented agentic framework that formulates vulnerability repair as an iterative, experience-driven process. \ourmodel{} combines three complementary memory layers, \textit{i.e.}, History-Fix, Security-Pattern, and Refinement-Trajectory memories, with a dynamic feedback-driven refinement loop. This design allows the agent to retrieve repository-specific repair conventions, apply reusable security defenses, and exploit prior ``failure-to-success'' trajectories to revise semantically invalid patches based on runtime evidence.
We evaluate \ourmodel{} on three representative repository-level vulnerability repair benchmarks: SEC-Bench, PatchEval (Python, Go, JavaScript), and the C++ subset of Multi-SWE-bench. \ourmodel{} achieves state-of-the-art resolution rates of 58.0\%, 58.2\%, and 30.58\%, respectively, outperforming strong general-purpose agents such as OpenHands and SWE-agent, as well as the specialized AVR tool InfCode-C++, while maintaining competitive repair cost. These results show that persistent, hierarchical repair memory can substantially improve the reliability of agentic vulnerability repair across diverse languages and repository settings. Our code and data are available at \url{https://figshare.com/s/f2bbe7f7c8a759339368}.

\end{abstract}



\received{20 February 2007}
\received[revised]{12 March 2009}
\received[accepted]{5 June 2009}

\maketitle

\section{Introduction}

Nowadays, unprecedented levels of software vulnerabilities critically threaten the global software supply chain with severe risks \cite{recordedfuture2025h1}.
Manual vulnerability repair is inefficient, high-cost, and thereby fundamentally unable to scale to the volume and velocity of modern vulnerability disclosures \cite{cyware2024challenges, vicarius2024remediation}.
As a result, the demand for scalable and reliable automated vulnerability repair (AVR) \cite{hu2025sokautomatedvulnerabilityrepair,li2025sokeffectiveautomatedvulnerability} approaches is steadily increasing, emerging as a vital research frontier.

To meet these demands, the paradigm of AVR has undergone substantial evolution.
Early approaches relied on templates~\cite{xuan2016nopol, liu2019tbar} or simple translation models~\cite{fu2022vulrepair}, and recently the field has progressed toward autonomous agents~\cite{bouzenia2024repairagent, zhang2024autocoderover, HOU2025107671,dong2025infcodecintentguidedsemanticretrieval} powered by Large Language Models (LLMs).
These agents utilize external toolsets to augment reasoning, aiming to handle complex vulnerabilities, such as those requiring cross-file dependency understanding, in a stable and efficient manner.
Despite these advancements, empirical evaluations~\cite{bui2024apr4vul, wei2025patcheval} indicate that existing approaches achieve only a relatively limited success rate in generating effective patches for real-world vulnerabilities.
More critically, as noted in analyses of tools like jKali~\cite{qi2015analysis}, many plausible patches introduce functional failures by simply deleting code or suppressing crashes without preserving semantics.
To mitigate this, recent literature has explored augmenting agents with static knowledge~\cite{liu2024crepair, DBLP} or iterative feedback loops~\cite{xia2024automated}. Yet, these methodologies fail to effectively synergize in repository-level environments.
Specifically, naive retrieval strategies frequently overlook the intricate software context and interface specifications inherent in large repositories. As empirically demonstrated in recent studies on RAG integration~\cite{shao2025llmscorrectlyintegratedsoftware, zhao2025recodeimprovingllmbasedcode}, this neglect causes retrieved generic patterns to clash with rigid project constraints, rendering the generated patches practically inapplicable.
Furthermore, while feedback-driven frameworks~\cite{xia2024automated, liu2025agentdebugsdynamicstateguided} can identify patch fails by test feedback, they lack the experiential memory to infer how to rectify it. Without the guidance of repair experiences, they treat feedback merely as rejection signals, leading to Cognitive Deficiency where agents oscillate between invalid solutions rather than generating a correct fix~\cite{wang2024openhands}.

In contrast, security experts do not analyze code context in isolation when resolving vulnerabilities. Instead, they draw on multiple layers of accumulated experience: \textit{project-specific knowledge} such as past fixes within the same repository, \textit{generalized security knowledge} such as well-known defensive patterns like boundary checks, and \textit{debugging experience} from prior failed-then-successful repair attempts. By combining these complementary knowledge sources, experts can reliably diagnose root causes and produce robust fixes through experience-guided refinement.

Motivated by the cognitive processes of security experts, we argue that, to achieve human-level reliability, autonomous agents require a \textit{memory-augmented architecture} capable of explicit experience retrieval and continuous learning.
Unlike current tools \cite{wang2024openhands,yang2024swe,Aider} that rely solely on the immediate context window, such an architecture maintains a knowledge base to store and recall verified solutions and debugging insights.
Although prior work has begun to augment LLM-based repair with retrieval, these components are often loosely coupled, treating feedback as transient signals for the current attempt rather than consolidating it into persistently retained and reusable refinement experience.
Consequently, these systems often repeat identical semantic errors across attempts or fail to generalize previously successful strategies to similar defects.
To bridge this gap, we introduce \ourmodel{}, a repository-level vulnerability repair framework designed to mimic the expert's cognitive hierarchy.
\ourmodel{} employs a collaborative multi-agent system supported by a structured \textbf{three-tier memory hierarchy}, which is constructed to address the three corresponding cognitive barriers
(\textit{i.e.}, lack of practical repair experience from humans, insufficient domain knowledge, and the absence of effective debugging insights.)

\begin{itemize}[leftmargin=*]
  \item \textbf{L1 History-Fix Memory} retrieves similar historical fixes from the same project to ensure patches adhere to project-specific conventions.
  \item \textbf{L2 Security-Pattern Memory} supplies generalized cross-project security knowledge (\textit{e.g.}, boundary checks) when local history is insufficient.
  \item \textbf{L3 Refinement-Trajectory Memory} stores ``failure-to-success'' trajectories from prior repair sessions, guiding iterative patch refinement based on dynamic feedback.
\end{itemize}
To evaluate the effectiveness of \ourmodel{}, we conduct comprehensive experiments on three rigorous benchmarks: \textbf{SEC-Bench}~\cite{lee2025secbench} (C/C++ memory safety vulnerabilities), \textbf{PatchEval}~\cite{wei2025patcheval} (multi-language real-world vulnerabilities), and the C++ subset of \textbf{Multi-SWE-bench}~\cite{zan2025multi} (large-scale repository issues).
Experimental results demonstrate that \ourmodel{} achieves resolution rates of
58.0\% on SEC-Bench and 58.2\% on PatchEval, outperforming the previously best-performing agents by more than 20 percentage points.
Furthermore, on the complex large-scale Multi-SWE-bench (C++), \ourmodel{} achieves a resolution rate of 30.6\%, exceeding the state-of-the-art specialize AVR tool \textbf{InfCode-C++} (25.60\%)~\cite{dong2025infcodecintentguidedsemanticretrieval} by 19\%, validating its superior capability in navigating intricate codebases and resolving challenging bugs.

In summary, this paper makes the following contributions:

\begin{itemize}[leftmargin=*]
  \item \textbf{Memory-Enhanced Framework:} We propose \ourmodel{}, the first repository-level AVR framework integrating a three-tier memory hierarchy (History-Fix, Security-Pattern, and Refinement-Trajectory) that enables experience-grounded patch generation.

    \item \textbf{Feedback-Driven Refinement:} We introduce a Locator-Patcher-Verifier workflow with a closed feedback loop that converts runtime failure signals into corrective cues for iterative patch refinement.

    \item \textbf{Extensive Evaluation:} We evaluate on SEC-Bench~\cite{lee2025secbench}, PatchEval~\cite{wei2025patcheval}, and Multi-SWE-bench~\cite{zan2025multi}, demonstrating state-of-the-art performance across four languages.
\end{itemize}
\section{Motivation}
We illustrate the core challenges of repository-level vulnerability repair with CVE-2023-0841, a critical \textit{integer overflow} in the \texttt{GPAC} multimedia framework (a detailed walkthrough is available in our replication package).

In \texttt{reframe\_mp3.c}, the allocation size \texttt{tag\_size + 10} wraps around when \texttt{tag\_size} approaches \texttt{UINT\_MAX}, causing \texttt{realloc} to return a tiny buffer. A subsequent \texttt{memcpy} then writes the original large payload into this undersized buffer, triggering a heap buffer overflow. OpenHands (DeepSeek-v3.2) fixes the integer overflow at the allocation site but misses the dependency with the downstream \texttt{memcpy}, leaving the write path unguarded. This exemplifies a common failure mode: without cross-context reasoning, agents apply local syntax corrections that fail to propagate security constraints along data-flow dependencies.

\ourmodel{} bridges this gap through its multi-tier memory. It retrieves a generalized L2 insight that allocation-site validation alone is insufficient, together with an L3 ``failure-to-success'' trajectory showing that pointer updates require post-allocation verification. Guided by this experience, \ourmodel{} injects a composite guard checking \texttt{ctx->id3\_buffer\_size + bytes\_to\_drop} before the write, preventing the unsafe path.

\section{Proposed Framework}

\subsection{Problem Formulation}\label{sec:define}

Formally, let $\mathcal{P}$ denote the target repository and $D_{desc}$ represent the vulnerability report. The system is also given a verification oracle $\langle \mathcal{T}, \tau_{vuln} \rangle$, comprising a regression test suite $\mathcal{T}$ and a vulnerability Proof-of-Concept (PoC) $\tau_{vuln}$.
The goal is to synthesize a patch $\delta$ yielding a patched version $\mathcal{P}' = \mathcal{P} \oplus \delta$ that satisfies two semantic constraints:

\noindent\textbf{(1) Functionality Preservation.} The patch must maintain existing functionality, ensuring no regressions on passing tests:
\begin{equation}
    \forall t \in \mathcal{T}: \mathcal{P}(t) = \textsc{Pass} \implies \mathcal{P}'(t) = \textsc{Pass}
\end{equation}

\noindent\textbf{(2) Vulnerability Mitigation.} The patch must resolve the defect captured by the PoC $\tau_{vuln}$ (where $\mathcal{P}(\tau_{vuln}) = \textsc{Fail}$):
\begin{equation}
    \mathcal{P}'(\tau_{vuln}) = \textsc{Pass}
\end{equation}

\noindent The vulnerability repair task is to find a $\delta$ satisfying Eq.~(1) and Eq.~(2), treating $\tau_{vuln}$ as the security target and $\mathcal{T}$ as the semantic boundary.

\begin{figure*}[t]
    \centering
    \setlength{\abovecaptionskip}{0.1cm}
    \includegraphics[width=\linewidth]{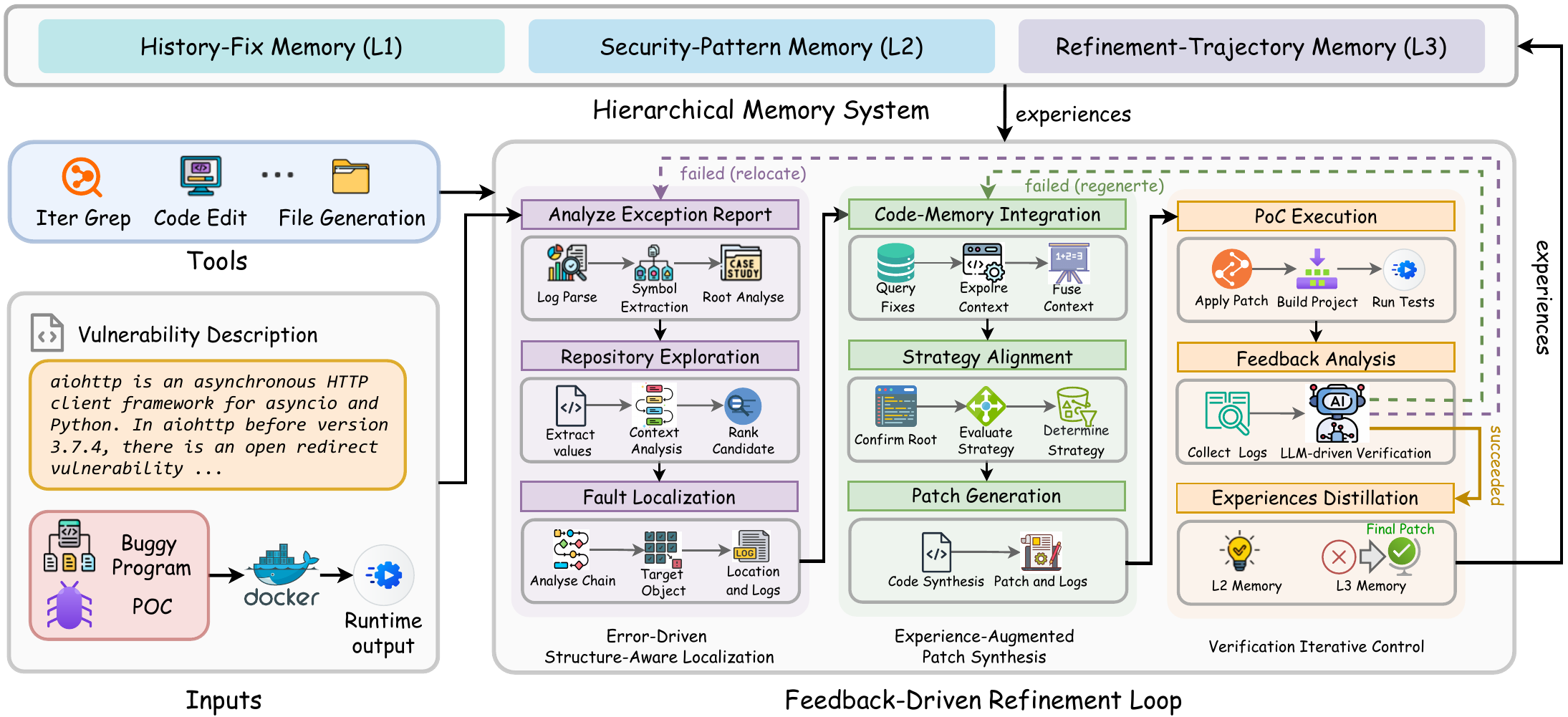}
    \caption{Overall architecture of \ourmodel{}.}
    \label{fig:overview}
    \vspace{-0.3cm}
\end{figure*}
\subsection{Framework Overview}
\label{sec:overview}
Figure~\ref{fig:overview} presents an overview of our proposed \ourmodel{}.
Our framework mimics the iterative cognitive process of human experts, aiming to deliver robust and reliable vulnerability fixes. Rather than relying on a isolated generation pass, \ourmodel{} orchestrates a dynamic repair lifecycle driven by a \textbf{Hierarchical Memory System} and a \textbf{Feedback-Driven Refinement Loop} to solve the task defined in Section~\ref{sec:define}.

\ourmodel{} first runs vulnerability  PoC $\tau_{vuln}$ to obtain runtime failure evidence (\textit{e.g.}, stack traces/sanitizer reports) and uses a structure-aware localizer tool (\texttt{Iter\_grep}) to produce a localization object $\mathcal{L}_{loc}$ (target file, suspicious line range, and a concise trace). Conditioned on $\mathcal{L}_{loc}$, the Patcher generates a \emph{candidate} patch $\delta$ guided by hierarchical memory retrieval. Finally, the Verifier accepts $\delta$ only if it passes the oracle $\langle \mathcal{T}, \tau_{vuln}\rangle$, otherwise it triggers \textsc{Regenerate} (revise $\delta$ at the same $\mathcal{L}_{loc}$) or \textsc{Relocate} (update $\mathcal{L}_{loc}$ with new runtime evidence), forming a closed feedback loop.
Details of memory construction/retrieval and the refinement loop (including the algorithmic design and I/O interface of \texttt{Iter\_grep}) are presented below.
Table~\ref{tab:tools} summarizes the tool interface provided to the agent for interacting with the repository $\mathcal{P}$, constructing $\mathcal{L}_{loc}$, and validating candidate patches $\delta$ against the oracle $\langle \mathcal{T}, \tau_{vuln}\rangle$.

\begin{table}[t]
\centering
\small
\setlength{\abovecaptionskip}{0.1cm}
\caption{
Tool interface of \ourmodel{}.
}
\label{tab:tools}
\resizebox{\linewidth}{!}{
\begin{tabular}
{p{0.2\linewidth}>{\raggedright\arraybackslash}p{0.35\linewidth}p{0.45\linewidth}}
\toprule
\multicolumn{1}{l}{\textbf{Category}}& \multicolumn{1}{c}{\textbf{Command}} & \multicolumn{1}{c}{\textbf{Documentation}} \\
\midrule

\textit{Localization}
& \textbf{Iter\_grep}\texttt{ <symbol>}
& Returns top-\texttt{k} (default \texttt{k=5}) ranked locations to form $\mathcal{L}_{loc}$: \texttt{\{file, line\_range\}}. \\

\cmidrule(lr){1-3}

\textit{File viewer}
& \textbf{view}\texttt{ <path>}
& Shows file content with line numbers; lists directory items (depth $\le$2). \\

\cmidrule(lr){1-3}

\textit{Search tools}
& \textbf{search}\texttt{ <pattern> <search\_path>}
& Regex search via \texttt{ripgrep}; returns matches with context (default matches number=5). \\

\cmidrule(lr){1-3}

\textit{File editing}
& \textbf{create}\texttt{ <path> <text>}
& Creates a new file (fails if \texttt{path} already exists). \\

\addlinespace
& \textbf{str\_replace}\texttt{ <path> <old> <new>}
& Replaces an exact, unique match of \texttt{old} with \texttt{new}. \\

\cmidrule(lr){1-3}

\textit{Command execution}
& \textbf{bash}\texttt{ <command> [restart]}
& Runs a command in a persistent shell session (\texttt{restart=true} resets). \\

\cmidrule(lr){1-3}

\textit{Verification}
& \textbf{check\_vul}
& Runs $\langle \mathcal{T}, \tau_{vuln}\rangle$ on $\mathcal{P}\oplus\delta$; returns verdict + error logs. \\

\cmidrule(lr){1-3}

\textit{Log compression}
& \textbf{log\_compress}
& Summarizes logs into a compact template for the next iteration. \\

\cmidrule(lr){1-3}

\textit{Task submission}
& \textbf{submit}
& Submits final result and extracted \texttt{git diff}. \\

\bottomrule
\end{tabular}
}
\vspace{-0.3cm}
\end{table}

\subsection{Hierarchical Memory Construction}
\label{sec:memory_construction}

To support the cognitive processes described in Section~\ref{sec:overview}, we formally define the data structures for the three-tier memory hierarchy.

\subsubsection{Level 1: History-Fix Memory ($M_{L1}$)}
The L1 memory is derived from CVEFixes~\cite{bhandari2021cvefixes},
comprising 14,063 verified vulnerability-fixing entries collected up to 2024. These historical vulnerabilities are mined from the National Vulnerability Database (NVD) \cite{NVD}, a public U.S. vulnerability repository maintained by NIST that aggregates CVE records and detailed metadata on known software vulnerabilities, many of which also appear in widely used vulnerability repair benchmarks (including our evaluation benchmarks). As a result, this corpus can effectively serve as a source of repository-relevant historical fix experience.
Formally, an L1 entry $e_{L1}$ is defined as a tuple:
\begin{equation}
   e_{L1} = \langle P_{proj}, K_{cwe}, L_{lang},I_{id}, D_{desc}, P_{fix} \rangle
\end{equation}
\begin{itemize}[leftmargin=*]
 \item $P_{roj}, K_{cwe}, L_{ang}$ (Retrieval Keys): The project identifier, vulnerability type, and target programming language, which collectively serve as the primary keys for precise experience retrieval.
\item $I_{id}$ (\textit{Instance ID}): A unique vulnerability identifier (such as \\ \texttt{njs.cve-2022-32414}), used to deduplicate entries and to filter out the target instance during retrieval (i.e., exclude $I_{id}(e)=I_{id}(q)$). We also use the CVE year and sequence (e.g., \texttt{2022-32414}) as a time key for temporal constraints.

    \item $D_{desc}$ (\textit{Description}): The detailed description of the vulnerability scenario.
    \item $P_{fix}$ (\textit{Solution}): The concrete repair patch containing the exact code modifications that resolved the vulnerability.
\end{itemize}
\noindent L2 and L3 memories share the same keys ($P_{proj}, K_{cwe}, L_{lang},
I_{id}, D_{desc}$) as L1. Both are collected in real-time during the repair process. Below we describe only their unique fields.

\subsubsection{Level 2: Security-Pattern Memory ($M_{L2}$)}
Each L2 entry $e_{L2}$ represents a successful repair instance, extending the shared keys with:
\begin{equation}
      e_{L2} = \langle \ldots, R_{rationale} \rangle
\end{equation}
where $R_{rationale}$ (\textit{Insight}) is a concise natural language description of why the patch worked.

\subsubsection{Level 3: Refinement-Trajectory Memory ($M_{L3}$)}
The L3 memory captures ``failure-to-success'' trajectories within a single repair session. Each entry extends the shared keys with:
\begin{equation}
    e_{L3} = \langle \ldots, P_{fail}, \Delta_{diff}, I_{insight} \rangle
\end{equation}
\begin{itemize}[leftmargin=*]
    \item $P_{fail}$ (\textit{Failure State}): The patch that failed validation.
    \item $\Delta_{diff}$ (\textit{Correction Delta}): The modification that transformed the failed state into a successful one.
    \item $I_{insight}$ (\textit{Transition Rule}): The learned rule for this transition.
\end{itemize}

The size of the evaluation benchmarks naturally limits the scale of the stored experiences (e.g., 116 for L2 and 86 for L3 on SEC-Bench). Since only successful fixes are accumulated, memory growth remains controlled in our current setting. To ensure scalability in long-running deployments across large-scale repositories, we adopt two lightweight management strategies: (1)~\textit{deduplication}, where entries whose descriptions and patches have a cosine similarity above 0.95 are merged to eliminate redundancy; and (2)~\textit{recency-weighted pruning}, where entries that have not been retrieved over a configurable window of recent tasks are periodically removed to keep the memory focused on actively useful knowledge.

\begin{figure}[t]
    \centering
    \setlength{\abovecaptionskip}{0.1cm}
      \includegraphics[width=\linewidth]{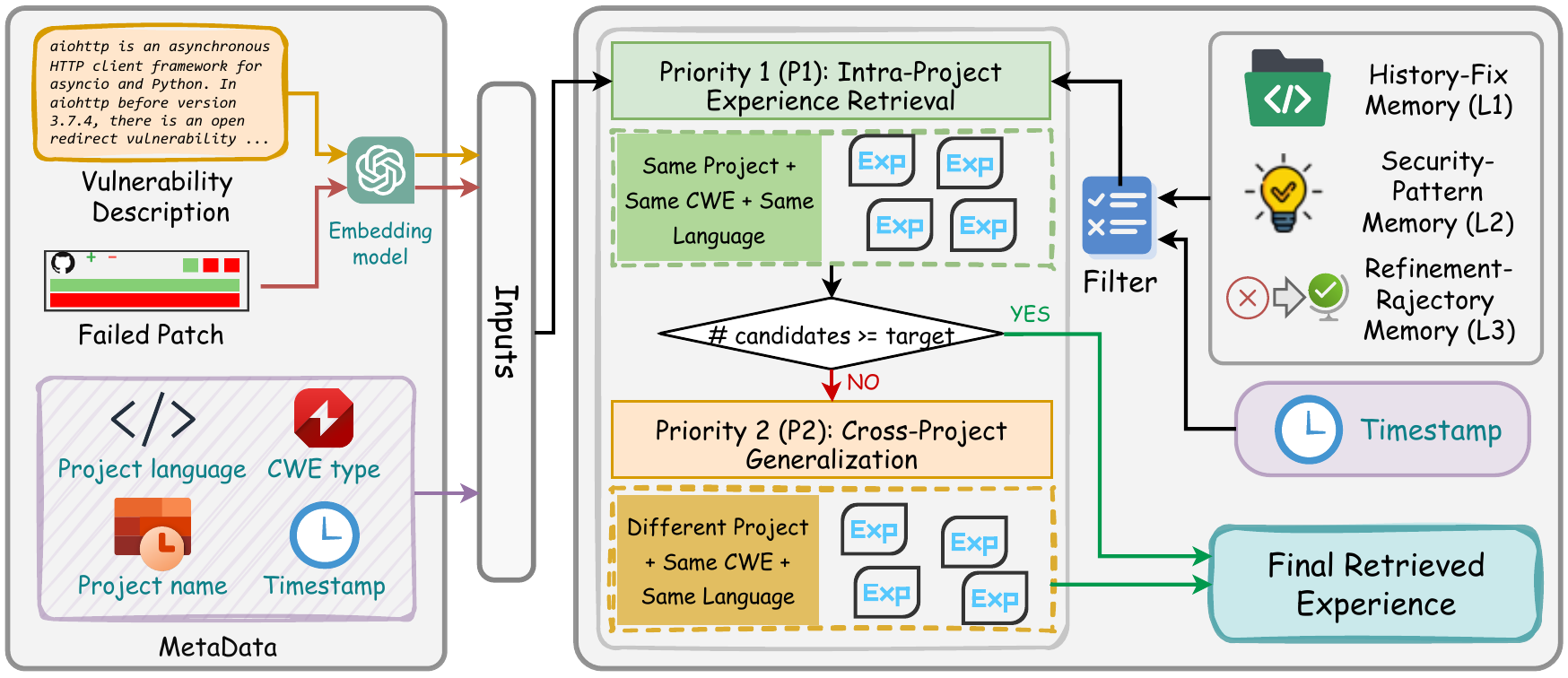}
    \caption{Workflow of our experience retrieval mechanism.
    }
    \label{fig:retrieval_process}
    \vspace{-0.3cm}
\end{figure}

\subsubsection{Experience Retrieval Mechanism}\label{sec:retrieval}
\ourmodel{} employs an experience retrieval mechanism to effectively leverage historical knowledge.
The classic ``plastic surgery hypothesis''~\cite{zhu2024empirical} posits that repair ingredients are predominantly found within the same project, ensuring stylistic consistency.
At the same time, to balance locality with generalization and to ensure robust retrieval under sparse project history,
we propose a dynamic \textbf{two-tier priority strategy (P1 and P2)}, as illustrated in Figure~\ref{fig:retrieval_process}.

\noindent\textbf{Priority 1 (P1): Intra-Project Experience Retrieval.
}
\textit{Criteria: Same Project + Same CWE + Same Language.}
We prioritize intra-project experiences to maximize stylistic and semantic consistency. This ensures that retrieved patches adhere to the project's established naming conventions, error-handling idioms, and architectural constraints, while strictly targeting the specific vulnerability category.

\noindent\textbf{Priority 2 (P2): Cross-Project Generalization.}
\textit{Criteria: Different Project + Same CWE + Same Language.}
When P1 yields fewer than $k$ candidates ($k{=}2$ in our setup), we extend retrieval to P2, which is particularly important for vulnerability categories whose fixes often rely on external defensive patterns absent in a local codebase~\cite{zhu2024empirical, make7040149}. By incorporating cross-project repair knowledge, P2 guarantees non-empty and semantically meaningful retrieval even when same-repository history is limited.

We quantify the semantic proximity between $D_{desc}$ and candidate descriptions. We utilize the \texttt{text-embedding-3-small} model~\cite{openai2024embedding} to encode descriptions into dense vectors and rank candidates by their cosine similarity. To prevent data leakage, we exclude candidates with the same instance ID as the query (i.e., $I_{id}(e)=I_{id}(q)$).
We further enforce a temporal constraint on intra-project retrieval by requiring $Timestamp(e) < Timestamp(q)$, where $Timestamp(\cdot)$ is derived from $I_{id}$ (year and sequence number).

\subsection{Feedback-Driven Refinement Loop}\label{sec:feedback}

\ourmodel{} operates through an iterative agentic loop anchored by tool-driven execution and feedback (Table~\ref{tab:tools}), augmented by a multi-tier memory system (\S\ref{sec:memory_construction}).
Concretely, we first obtain dynamic failure evidence by invoking \textbf{\texttt{check\_vul}}, which executes the oracle $\langle \mathcal{T}, \tau_{vuln}\rangle$ on the current repository state and returns structured runtime evidence $R_{dyn}$ such as stack traces/sanitizer reports and failing tests.
For each failed attempt, we additionally invoke \textbf{\texttt{log\_compress}} to summarize raw logs into a compact template, which is fed back to the next iteration to reduce context noise.

\paragraph{Phase 1: Error-Driven Structure-Aware Localization}
The Locator agent ingests $D_{desc}$ and $R_{dyn}$, supplemented
by project-specific priors from L1/L2 memories. To overcome the limitations of keyword-based search (existing agents rely on
\texttt{grep} and return many unranked matches including
comments and unused code), we implement \textbf{\texttt{Iter\_grep}},
a structure-aware tool that returns the top-5 code locations most relevant to the error stack trace in $R_{dyn}$.
Given a symbol name (e.g., variable names from $D_{desc}$
or function names from $R_{dyn}$), \texttt{Iter\_grep} parses the repository AST to locate all definitions and use sites, then ranks them by proximity to the error location in $R_{dyn}$. It prioritizes matches within the files present in the $R_{dyn}$ call stack, ranking frames closer to the crash site higher by their logical distance to the crash point. The \texttt{Iter\_grep} output is a structured localization object $\langle f,\ell \rangle$, where $f$ is the ranked file path, $\ell$ is the critical line. This object serves as the deterministic target for subsequent patch synthesis.
For example, given a heap-buffer-overflow in \texttt{utils.c:45} inside function \texttt{safe\_copy(src, len)}, querying \texttt{Iter\_grep(len)} returns:
\begin{itemize}[leftmargin=*, noitemsep, topsep=0pt]
    \item Rank 1: \texttt{utils.c:45} (Crash site: \texttt{memcpy(dst, src, len)}).
    \item Rank 2: \texttt{utils.c:40} (Function signature definition).
    \item Rank 3: \texttt{main.c:102} (Caller site: \texttt{safe\_copy(buf, user\_input\_size)}).
\end{itemize}

\noindent Finally, this phase produces a structured localization object $\mathcal{L}_{loc}$, encapsulating the file path, critical line number, and localization reasons.

\paragraph{Phase 2: Experience-Augmented Patch Synthesis}
Conditioned on $\mathcal{L}_{loc}$ from Phase 1, \ourmodel{} proceeds to the Patcher Agent. Similar to the Locator, this agent is initialized with project-specific patterns from L1 and L2 memories to ensure the generated code adheres to existing stylistic and security conventions. It is worth noting that \textbf{the L3 refinement-trajectory memory is activated only after a failed verification cycle}.
Given a failed candidate $\delta$, we embed the failed patch and retrieve the most similar $P_{fail}$ entries from L3 under the same priority strategy.
The retrieved ``failure$\rightarrow$success'' deltas provide explicit corrective guidance, which the Patcher uses to revise $\delta$ in the subsequent iteration.

\paragraph{Phase 3: Verification Iterative Control}
The Verifier applies the candidate patch $\delta$ produced in Phase 2 to the repository and executes the verification oracle $\langle \mathcal{T}, \tau_{vuln} \rangle$ (via \textbf{\texttt{check\_vul}}) to validate both functionality preservation and vulnerability mitigation. Based on the execution feedback, it triggers a tri-state transition:
\begin{itemize}[leftmargin=*]
    \item \textbf{\textsc{Success}}: Terminates the lifecycle. The Verifier extracts the successful repair session to update L2, and if applicable, captures the ``failed $\to$ success'' transition to update L3.
    \item \textbf{\textsc{Relocate}}: Backtracks to Phase 1 if the vulnerability persists, indicating an incorrect root cause diagnosis.
    \item \textbf{\textsc{Regenerate}}: Loops back to Phase 2 if the patch fixes the vulnerability but introduces functional regressions.
\end{itemize}

To keep the next iteration focused and reduce token consumption, we perform \textit{context compression}:  the Verifier summarizes the failed trajectory from raw tool logs using \textbf{\texttt{log\_compress}} into a fixed template containing only (i) visited files/line ranges, (ii) applied diff hunks, and (iii) verification failure log (including sanitizer traces, or compilation errors). We then roll back the environment for the subsequent refinement cycle.

\section{Experimental Setup}

\subsection{Research Questions}
We evaluate the effectiveness of \ourmodel{} by comparing it with the state-of-the-art baselines and focus on the following research questions:

\begin{itemize}[leftmargin=*]
    \item \textbf{RQ1: Overall Performance} - How does \ourmodel{} compare to state-of-the-art automated vulnerability repair techniques and general-purpose LLM agents in fixing real-world vulnerabilities?
    \item \textbf{RQ2: Ablation Study} - What is the contribution of each memory component and the feedback-driven refinement loop to the overall repair performance?

    \item \textbf{RQ3: Generalizability Analysis} - Can \ourmodel{} generalize effectively across different programming languages (Python, Go and JavaScript) and varying levels of repository complexity?

\end{itemize}

\subsection{Baselines}

To validate the performance of \ourmodel{}, we compare it against two categories of state-of-the-art approaches. First, for general-purpose autonomous software engineering, we select representative agents including \textbf{OpenHands \cite{wang2024openhands}}, \textbf{SWE-agent \cite{yang2024swe}}, \textbf{Aider \cite{Aider}}, and \textbf{Agentless}. Specifically, OpenHands is evaluated using both \texttt{DeepSeek-v3.2} and \texttt{Claude-3.7-Sonnet} backbones, while SWE-agent utilizes \texttt{Claude-3.7-Sonnet} and \texttt{Gemini-2.5}. These frameworks represent the current state-of-the-art in agentic planning and code editing strategies, operating without domain-specific customization. we exclude prior learning-based AVR approaches (including both DL models~\cite{fu2022vulrepair, jiang2021cure} and non-agentic LLM applications~\cite{pearce2023examining, xia2023plastic,zhang2024prompt}) as they are predominantly restricted to \textit{function-level} synthesis with pre-isolated context and static evaluation such as CodeBLEU \cite{codebleu}. These are incompatible with our \textit{end-to-end repository-level} setting, which demands autonomous localization and rigorous dynamic security verification.

Furthermore, on the Multi-SWE-bench (C++ subset), we benchmark against \textbf{InfCode-C++}~\cite{dong2025infcodecintentguidedsemanticretrieval}. We prioritize this comparison as InfCode-C++ epitomizes the structural-retrieval paradigm, employing intent-guided search and deterministic AST-based navigation to resolve C++ complexities and achieve state-of-the-art performance. This comparison directly evaluates whether \ourmodel{}'s experience-driven memory hierarchy surpasses a system deeply engineered with language-specific static analysis. Additionally, while recent dynamic agents like VulDebugger~\cite{liu2025agentdebugsdynamicstateguided} show promise, they are excluded due to reproducibility constraints regarding their closed-source datasets. Besides, to verify generalizability beyond C++, we additionally incorporate PatchEval for multi-language evaluation (Go, Python, JavaScript).

\subsection{Benchmarks and Metrics}
To comprehensively evaluate the effectiveness of \ourmodel{} across different vulnerability types, programming languages, and repository scales, we employ three distinct benchmarks:
    (1) \textbf{SEC-Bench}~\cite{lee2025secbench}, a rigorous framework designed to validate broad security capabilities, specifically targeting C/C++ memory safety tasks by utilizing sanitizer feedback to evaluate agents on vulnerability patching;
    (2) \textbf{PatchEval}~\cite{wei2025patcheval}, a specialized benchmark for real world vulnerabilities including \textbf{Go, Python, and JavaScript}, which employs rigorous security and regression tests to verify that fixes effectively mitigate vulnerabilities without breaking functionality; and
    (3) the C++ subset of \textbf{Multi-SWE-bench}~\cite{zan2025multi}, representing the complex large-scale repository-level issue resolution, where we benchmark against \textbf{InfCode-C++}~\cite{dong2025infcodecintentguidedsemanticretrieval}. We restrict this comparison to Multi-SWE-bench (C++) because InfCode-C++ is C++-specific and its runnable implementation is not publicly available, making it infeasible to reproduce it under the SEC-Bench.

We employ the following metrics to evaluate both the vulnerability repair performance as well as the cost:

\begin{itemize}[leftmargin=*]
    \item \textbf{Success Rate (\% Resolved):} The percentage of tasks where the generated patch successfully passes both the vulnerability reproduction test ($\tau_{vuln}$) and the existing regression test suite ($\mathcal{T}$).

    \item \textbf{Location Accuracy:} We evaluate localization at file levels. Following \citet{yang2024swe}, we consider the location correct if the locations modified by the generated patch include all locations in the ground truth patch.

    \item \textbf{Cost:} The average monetary cost per task (in USD), calculated based on the token usage.
\end{itemize}

\subsection{Implementation Details}

We primarily utilize \textbf{DeepSeek-v3.2} and \textbf{DeepSeek-v3} as the foundational reasoning backbones for \ourmodel{}. These models were selected for their balance between performance and computational cost. For hyperparameter configuration, we follow DeepSeek's official recommendation for coding/math tasks and set the temperature to $0.0$ to encourage deterministic patch generation \cite{guo2024deepseek}.
We cap the refinement loop at 3 failed patch attempts based on a small-scale pilot study, as most successful cases converge within the first few iterations while additional retries mainly increase cost.

Regarding the baselines, we also reproduce OpenHands (denoted as \textbf{OpenHands*}). Specifically, to ensure a rigorous and fair comparison with our framework,
we augment the standard OpenHands agent with a specialized validation tool function named \texttt{check\_vul} and modify the system prompt to explicitly instruct the agent to invoke this tool whenever it is confident that a fix has been achieved. Upon invocation, \texttt{check\_vul} executes the evaluation script on the current state of the repository and returns the execution results to the agent, thereby enabling a validation feedback loop.

\section{Results and Analysis}\label{sec:results}

\subsection{RQ1: Overall Performance}\label{sec:rq1}

To answer this RQ, we evaluate \ourmodel{} against state-of-the-art baselines on SEC-Bench and Multi-SWE-bench (C++ subset), both of which focus on C++ repositories.
\begin{table}[t]
    \centering
    \small
    \setlength{\abovecaptionskip}{0.1cm}
    \caption{Performance comparison on SEC-Bench and Multi-SWE-bench (C++).}
    \label{tab:performance_breakdown}
    \resizebox{\linewidth}{!}{
    \begin{tabular}{llccc}
        \toprule
        \textbf{Tool} & \textbf{LLM} & \textbf{\% Res.} & \textbf{Cost (\$/Task)} & \textbf{\% Cor. Loc.} \\
        \midrule
        \multicolumn{5}{l}{\cellcolor{gray!10}\textbf{\textit{SEC-Bench}}} \\
        \midrule
        \textbf{\ourmodel{}} & \textbf{DeepSeek-v3.2} & \textbf{58.00\%} & 0.26 & 60.00\% \\
        \hdashline[2pt/2pt]
        OpenHands* & DeepSeek-v3.2 & 38.50\% & 0.18 & 63.00\% \\
        OpenHands & Claude-3.7 & 34.00\% & 0.61 & 65.00\% \\
        SWE-agent & Claude-3.7 & 31.50\% & 1.29 & 67.50\% \\
        Aider & Claude-3.7 & 23.50\% & 0.44 & 46.00\% \\
        \midrule
        \multicolumn{5}{l}{\cellcolor{gray!10}\textbf{\textit{Multi-SWE-bench (C++)}}} \\
        \midrule
        \textbf{\ourmodel{}} & \textbf{DeepSeek-v3.2} & \textbf{30.58\%} & 0.32 & 53.72\% \\
        \ourmodel{} & DeepSeek-v3 & 15.08\% & - & 38.89\% \\
        \hdashline[2pt/2pt]
        InfCode-C++ \cite{dong2025infcodecintentguidedsemanticretrieval} & GPT-5 & 25.60\% & - & 55.10\% \\
        InfCode-C++ \cite{dong2025infcodecintentguidedsemanticretrieval} & DeepSeek-v3 & 13.20\% & - & - \\
        MOpenHands & Claude-3.7 & 14.70\% & 0.22 & 39.53\% \\
        MSWE-agent & Claude-3.7 & 11.60\% & 0.18 & 23.26\% \\
        MAgentless & Claude-3.7 & 3.90\% & 0.43 & 17.05\% \\
        \bottomrule
    \end{tabular}
    }
    \vspace{-0.3cm}
\end{table}
\subsubsection{Overall Performance Comparison (RQ1.1)}\label{sec:rq1-1}
As shown in Table~\ref{tab:performance_breakdown}, \ourmodel{} achieves state-of-the-art performance, substantially outperforming both general-purpose agents and specialized tools.
Specifically, on SEC-Bench, \ourmodel{} resolves 58\% of issues, surpassing OpenHands* by nearly 20 percentage points (58.0\% vs.\ 38.5\%). Crucially, both systems use the identical \texttt{DeepSeek-v3.2} backbone and share the same \texttt{check\_vul} verification tool, confirming that the performance improvement is attributable to our memory-guided architecture rather than differences in LLM capability or tool access.
In terms of localization, \ourmodel{} achieves slightly lower localization accuracy than several baselines, this difference largely reflects a limitation of the localization metric rather than the system's true effectiveness. Specifically, \ourmodel{} exhibits a substantially higher conversion rate from localization to successful repair. In contrast, baseline methods frequently fail to produce valid repairs even when the bug location is correctly identified (see \S\ref{sec:rq1-2} for a detailed analysis).
Regarding cost efficiency, \ourmodel{} maintains a highly competitive average cost of \$0.26/task, comparable to OpenHands* (\$0.18) and significantly lower than SWE-agent (\$1.29).

On the more challenging Multi-SWE-Bench (C++), \ourmodel{} achieves 30.58\%, outperforming InfCode-C++ (25.6\% with GPT-5) despite using a weaker backbone. When controlling for the same DeepSeek-V3 backend, \ourmodel{} (15.08\%) also exceeds InfCode-C++ (13.20\%) and general agents such as MOpenHands (14.70\%) and MSWE-agent (11.60\%).
\vspace{3mm}
\begin{custommdframed}
\textbf{Answer to RQ1.1:} \ourmodel{} establishes a new state-of-the-art in vulnerability repair, substantially outperforming both general-purpose agents and specialized tools on SEC-Bench and Multi-SWE-bench. Notably, \ourmodel{} surpasses these specialized baselines while maintaining superior cost-efficiency, demonstrating that a memory-augmented architecture is critical for resolving complex repository-level vulnerabilities.
\end{custommdframed}

\subsubsection{Localization-to-Repair Gap (RQ1.2)}\label{sec:rq1-2}
\begin{figure*}[t]
  \centering
  \setlength{\abovecaptionskip}{0.1cm}
  \begin{subfigure}{0.45\linewidth}
    \centering
    \includegraphics[width=\linewidth]{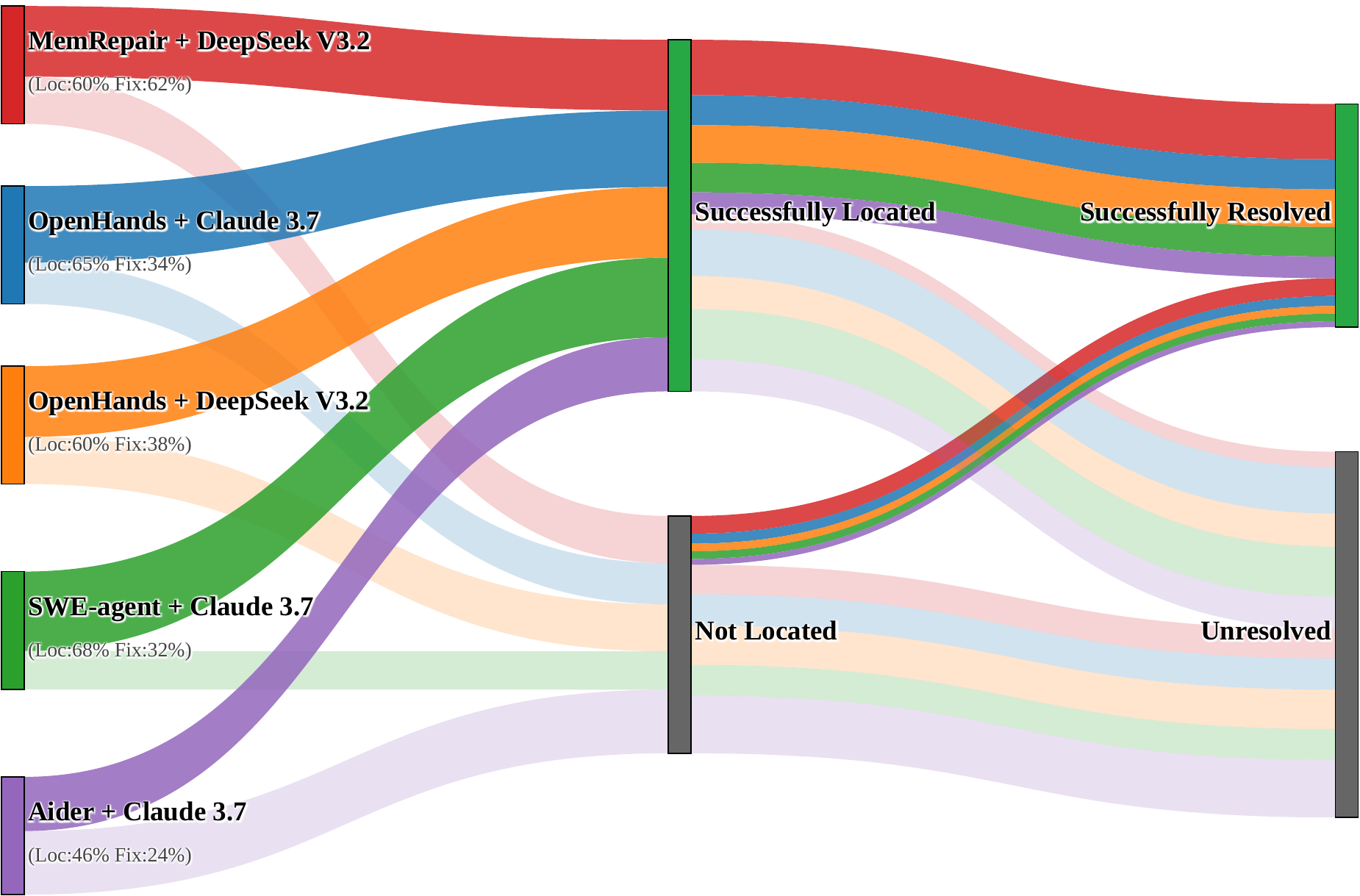}
    \caption{Results on SecBench.} 
    \label{fig:secbench}
  \end{subfigure}
  \hfill 
  \begin{subfigure}{0.44\linewidth}
    \centering
    \includegraphics[width=\linewidth]{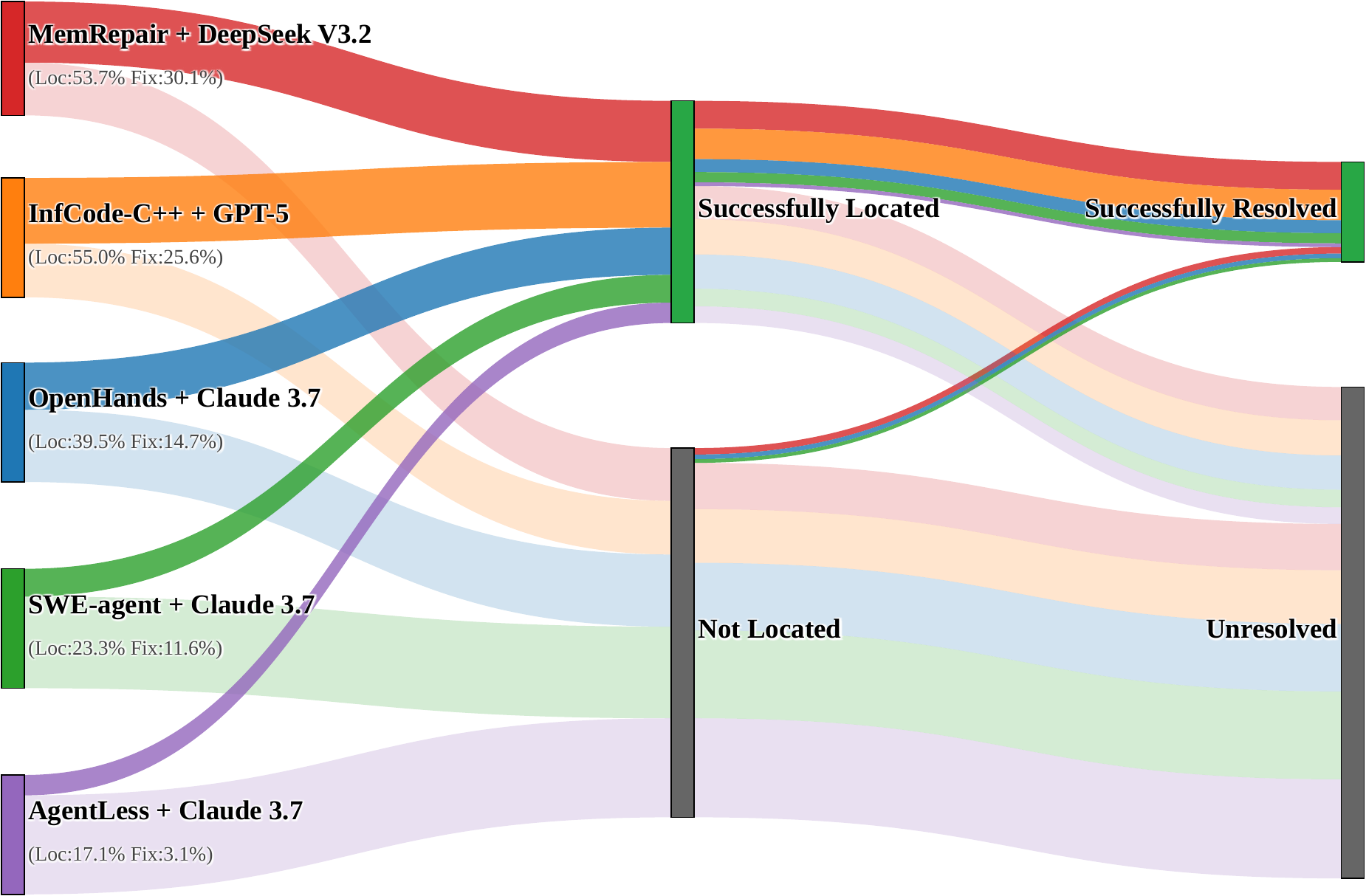}
    \caption{Results on Multi-SWE-Bench (C++).}
    \label{fig:muticpp}
  \end{subfigure}
  \caption{Issue localization-to-repair flow on SecBench and Multi-SWE-Bench (C++).}
  \label{fig:comparison}
  \vspace{-0.3cm}
\end{figure*}

To further analyze the gap between vulnerability localization and successful repair, we visualize the localization-to-repair flow using the Sankey diagram in Figure~\ref{fig:comparison}.
On SEC-Bench, baselines such as SWE-agent exhibit a seemingly superior localization rate of 67.5\% (135/200) compared to \ourmodel{} (60.0\%). However, high localization accuracy does not necessarily translate into more effective repair. Since the localization metric considers a prediction successful as long as the modified files \textit{include} the ground-truth vulnerable locations, a high score can sometimes reflect an unfocused editing strategy where the agent applies extensive edits across multiple files. While such aggressive edits increase the statistical likelihood of covering the buggy lines, they substantially raise the risk of introducing functional regressions or inconsistent fixes. The Sankey flow corroborates this observation:
among the vulnerabilities successfully localized by SWE-agent, 63\% (85 out of 135) ultimately fail to be repaired.

In contrast, \ourmodel{} demonstrates a higher conversion rate from localization to resolution. Out of the 120 instances it successfully localizes, 86 are correctly repaired, yielding a conversion rate of 71.6\%. By leveraging memory experiences, \ourmodel{} is guided toward generate precise, surgical fixes rather than broad and unfocused modifications.
This advantage becomes even more pronounced on the large-scale Multi-SWE-Bench (C++) benchmark. As illustrated in Figure~\ref{fig:muticpp}, OpenHands exhibits a low localization-to-repair conversion rate: it resolves only 27.5\% of the localized issues, leaving 72.5\% unresolved. \ourmodel{} maintains a robust conversion rate (45.5\%), demonstrating its capability to handle the noise and dependency complexity of large-scale software.

\vspace{3mm}
\begin{custommdframed}

\textbf{Answer to RQ1.2:} High localization accuracy does not guarantee repair success. Baselines exhibit a significant ``Localization-to-Repair'' gap, failing to resolve over 60\% of the bugs they successfully localize. In contrast, \ourmodel{} demonstrates a superior conversion rate from localization to valid repair, validating the effectiveness of memory-guided surgical repair.

\end{custommdframed}

\subsubsection{Solution Uniqueness and Overlap (RQ1.3)}\label{sec:rq1-3}
\begin{figure}[t]
    \setlength{\abovecaptionskip}{0.1cm}
    \centering
    \includegraphics[width=0.7\linewidth]{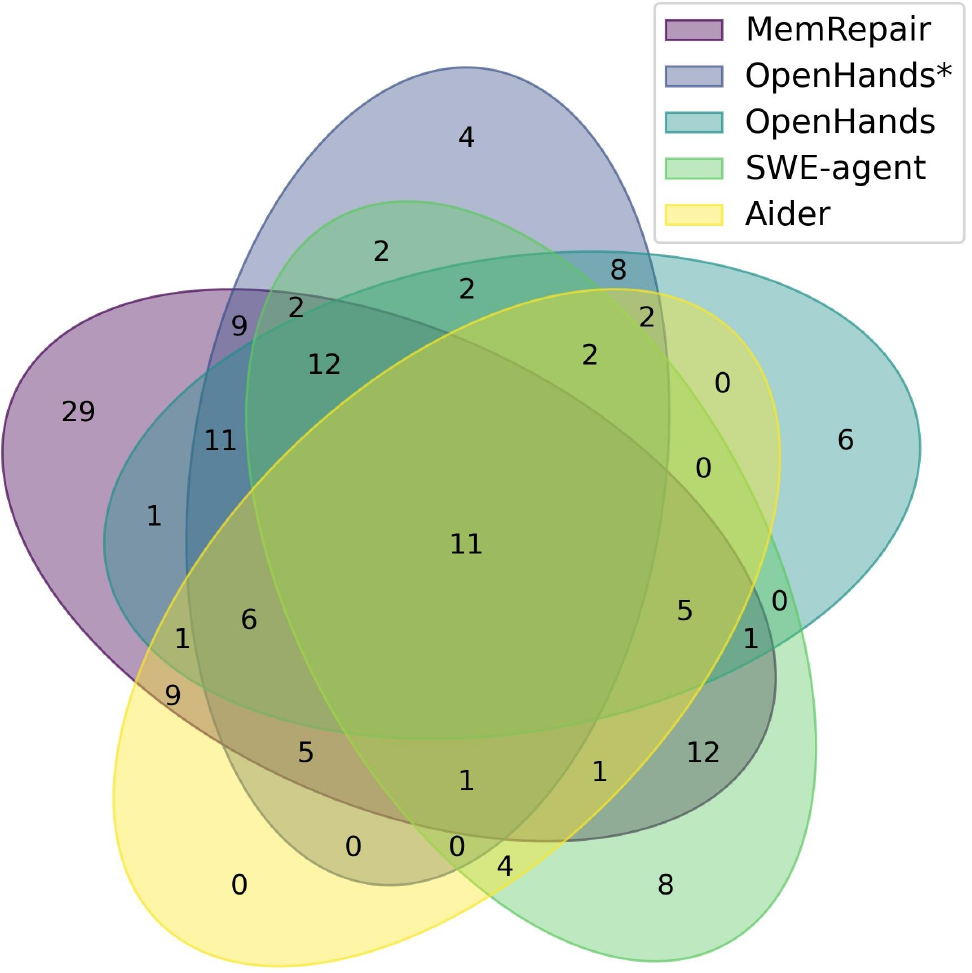}
    \caption{Venn diagram of the number pf resolved issues on SEC-Bench.}
    \label{fig:ven}
    \vspace{-0.3cm}
\end{figure}

Finally, we analyze whether \ourmodel{} simply replicates existing capabilities or addresses previously unsolvable problems. Figure~\ref{fig:ven} illustrates the intersection of resolved issues on SEC-Bench for \ourmodel{} and leading baselines.
Collectively, the union of all evaluated tools resolved \textbf{154} distinct vulnerabilities. \ourmodel{} alone covers \textbf{116} of these (approx. 75.3\%), demonstrating its role as the dominant contributor to the combined success pool.
More importantly, as seen in the figure, \ourmodel{} uniquely resolves \textbf{29} issues (25\% of its total successes).

In sharp contrast, the baselines exhibit much narrower unique capabilities, with SWE-agent and OpenHands uniquely resolving only 8 and 6 issues, respectively. This indicates that \ourmodel{} does not merely improve upon general agents by a small margin but possesses a distinct capability profile capable of solving a class of problems that are currently out of reach for general-purpose LLMs.

\vspace{3mm}
\begin{custommdframed}
\textbf{Answer to RQ1.3:} \ourmodel{} resolves 29 difficult vulnerabilities that no other agents could fix—more than \textbf{3$\times$} the unique contribution of the next best baseline (SWE-agent: 8). This confirms that \ourmodel{} offers distinct capabilities, effectively resolving intractable vulnerabilities that remain out of reach for general-purpose baselines.

\end{custommdframed}

\subsection{RQ2: Ablation Study}\label{sec:rq2}

To investigate the individual contribution of each memory component, \textit{i.e.}, L1/L2/L3 memory, we conducted an ablation study on SEC-Bench.
We compared the full \ourmodel{} framework against configurations where only a single memory tier was activated. Table~\ref{tab:ablation_study_detailed} details the success rates and the breakdown of localization-repair outcomes.

\begin{table*}[t] 
    \centering
    \setlength{\abovecaptionskip}{0.1cm}
    \caption{Ablation study results on SecBench. $L_{\checkmark/\times}$ denotes successful/failed localization, while $F_{\checkmark/\times}$ denotes successful/failed fixing.}
    \label{tab:ablation_study_detailed}
    \begin{tabular}{l c c c c c c c}
        \toprule
        & & & & \multicolumn{4}{c}{\textbf{Location Breakdown}} \\
        \cmidrule(lr){5-8}
        \textbf{Configuration} & \textbf{Base Model} & \textbf{\% Resolved} & \textbf{Cost (\$/Task)} &
        \boldmath$L_{\checkmark} F_{\checkmark}$ & \boldmath$L_{\checkmark} F_{\times}$ & \boldmath$L_{\times} F_{\checkmark}$ & \boldmath$L_{\times} F_{\times}$ \\
        \midrule
        \textbf{\ourmodel{} (Full: L1+L2+L3)} & DeepSeek-v3.2 & \textbf{58.0\%} & 0.26 & 86 & 34 & 30 & 50 \\
        \midrule
        \ourmodel{} (L3: Only) & DeepSeek-v3.2 & \underline{50.5}\% & 0.19 & 66 & 30 & 35 & 69 \\
      \ourmodel{} (L2 Only) & DeepSeek-v3.2 & 48.0\% & 0.20 & 77 & 38 & 19 & 66 \\
        \ourmodel{} (L1 Only) & DeepSeek-v3.2 & 42.0\% & 0.20 & 63 &
        41 & 21 & 75 \\
        \ourmodel{} (\textbf{Loop Only}) & DeepSeek-v3.2 & 36.0\% & 0.21 & 57 &
        43 &15 & 85 \\

        \midrule
       OpenHands* & DeepSeek-v3.2 & 38.5\% & 0.18 & 65 & 61 & 12 & 62 \\
        \bottomrule
    \end{tabular}
\end{table*}

\noindent\textbf{Impact of Refinement-Trajectory Memory (L3).}
The results highlight that \textit{L3 Refinement-Trajectory Memory} is the most critical driver of repair success. When operating with L3 alone, the system achieves a success rate of 50.5\%, significantly outperforming the OpenHands baseline (38.5\%).
More importantly, L3 drastically improves the \textit{quality} of the generated patches. As shown in the $L_{\checkmark}F_{\times}$ column (correct location, failed fixing), the baseline OpenHands fails to resolve 66 cases even after correctly locating them. In contrast, the L3-only configuration reduces this failure count to 30, and the full model further suppresses it to 26. This reduction demonstrates that the ``failure-to-success'' trajectories stored in L3 effectively guide the agent to rectify semantic errors that would otherwise lead to invalid patches, thereby bridging the localization-repair gap.

\noindent\textbf{Impact of Security-Pattern (L2) and History-Fix (L1) Memories.}
Both L1 and L2 memories contribute to performance improvements over the baseline, but with varying degrees of impact.
\textit{L2 Security-Pattern Memory} alone achieves a 48\% success rate, surpassing \textit{L1 History-Fix Memory} (42.0\%).
This result highlights that treating historical fixes as reusable code snippets, rather than as sources of abstract security knowledge, offers limited benefits for vulnerability repair.
For security vulnerabilities, generalized abstract patterns (\textit{e.g.}, ``check bounds before write'') are more transferable and effective than raw, project-specific code snippets retrieved by L1.

\noindent\textbf{Impact of Feedback Loop (\ourmodel{} \textbf{Loop Only}).}
The \textit{Loop Only} variant disables all memory retrieval (L1--L3) and retains only the feedback-driven control logic in the Verifier. Importantly, the Verifier in this setting provides \emph{only coarse-grained signals}, \textit{i.e.}, \textsc{Relocate} or \textsc{Regenerate}, to indicate whether the next attempt should change the localization target or revise the patch at the same location, without supplying any additional experience cues or refinement strategies. As shown in Table~\ref{tab:ablation_study_detailed}, \textit{Loop Only} variant achieves a success rate of 36.0\%, which is lower than OpenHands* (38.5\%) and substantially worse than any memory-enabled configuration (42.0\%--50.5\%). This result indicates that without memory-based guidance, the agent struggles to correct semantic mistakes, even after locating the correct region. This is reflected in the high $L_{\checkmark}F_{\times}$ count (43), where the system fails to generate a valid fix despite identifying the correct location.

\noindent\textbf{Effect of Hierarchical Memory.}
The Full \ourmodel{} configuration yields the highest success rate of 58.0\%, which is superior to any single-component setup. This indicates a synergistic effect: L1 and L2 provide the necessary context and search space reduction during the initial generation, while L3 acts as a semantic filter and refinement engine during the feedback loop.
Furthermore, the cost analysis reveals that this performance gain is achieved economically. The full model incurs only a marginal cost increase (\$0.26/task) compared to the baseline (\$0.18/task), proving that augmenting LLMs with a structured memory hierarchy is a cost-effective strategy for enhancing automated vulnerability repair.

\vspace{3mm}
\begin{custommdframed}

\textbf{Answer to RQ2:} The ablation study reveals that Refinement-Trajectory Memory (L3) stands out as the most significant individual contributor to repair success. While generalized security patterns (L2) prove more effective than raw historical snippets (L1), the full hierarchical synergy yields the highest performance. Besides, the ``Loop Only'' variant underperforms the baseline, proving that coarse feedback signals are insufficient without memory context.

\end{custommdframed}

\subsection{RQ3: Generalizability Analysis}\label{sec:rq3}

In this RQ, we investigate whether \ourmodel{} generalizes effectively across programming languages. We evaluate \ourmodel{} on PatchEval~\cite{wei2025patcheval}, which contains real-world vulnerabilities spanning \textbf{Python}, \textbf{JavaScript}, and \textbf{Go}, and validates fixes using a strict oracle $\langle \mathcal{T}, \tau_{vuln}\rangle$ (security reproduction + regression tests). To adapt to the benchmark, we only use generic text-based search utilities (standard \texttt{grep} or \texttt{ripgrep}), excluding the language-dependent AST analyzer (\texttt{Iter\_grep}) used in the C++ experiments.
As shown in Table~\ref{tab:rq3_patcheval_detail}, \ourmodel{} achieves an overall repair success rate of 58.2\% on PatchEval across three languages, comparable to its performance on SEC-Bench, indicating that the proposed memory-guided synthesis and feedback-driven refinement are not confined to C/C++ project scenarios but transfer to diverse languages (The total number of this benchmark is 230, but five docker images failed to build locally, so they were excluded.).
\begin{table}[t]
    \centering
    \small
    \setlength{\abovecaptionskip}{0.1cm}
    \caption{Generalizability results on PatchEval (Python/JavaScript/Go).}
    \label{tab:rq3_patcheval_detail}
    \begin{tabular}{lccc}
        \toprule
        \textbf{Setting} & \textbf{\# Cases} & \textbf{\% Located} & \textbf{\% Resolved} \\
        \midrule
        \multicolumn{4}{l}{\cellcolor{gray!10}\textbf{\textit{\ourmodel{} (DeepSeek-v3.2)}}} \\
        \midrule
        Python     & 65  & 70.8\% & 46.2\% \\
        JavaScript & 75  & 88.0\% & 74.7\% \\
        Go         & 85  & 72.9\% & 53.0\% \\
        \midrule
        \textbf{Total} & \textbf{225} & \textbf{77.3\%} & \textbf{58.2\%} \\
        \midrule
        \multicolumn{4}{l}{\cellcolor{gray!10}\textbf{\textit{Baselines reported in PatchEval}}} \\
        \midrule
        SWE-agent (Gemini-2.5 pro)  & -- & -- & 37.80\% \\
        OpenHands (Gemini-2.5 pro)  & -- & -- & 35.60\% \\
        \bottomrule
    \end{tabular}
    \vspace{-0.3cm}
\end{table}

Performance varies across languages: JavaScript achieves the highest repair rate (74.7\%) with an 81.8\% localization-to-repair conversion, while Python is the most challenging (46.2\%, 60.9\% conversion). A notable portion of tasks fall into the $L_{\times}F_{\checkmark}$ category, indicating that \ourmodel{} produces behaviorally equivalent fixes at alternative locations (\textit{e.g.}, upstream sanitization) rather than modifying the exact ground-truth lines.

Besides, \ourmodel{} substantially outperforms strong baselines reported by the benchmark: SWE-agent (Gemini-2.5 pro) at 37.80\% and OpenHands (Gemini-2.5 pro) at 35.60\% (Detailed location breakdown is omitted for these baselines as their specific generated patches were not open-sourced by PatchEval~\cite{wei2025patcheval}). Moreover, \ourmodel{} is notably more cost-efficient, achieving a lower average cost of 0.25 compared to 2.39 and 4.64 per task for SWE-agent and OpenHands, respectively.

\vspace{3mm}
\begin{custommdframed}

\textbf{Answer to RQ3:} \ourmodel{} demonstrates robust generalizability across diverse programming languages, substantially outperforming strong baselines while reducing operational costs by over 3$\times$. The consistent performance across Python, JavaScript, and Go confirms that the proposed memory-driven architecture is language-agnostic and effectively transfers to varying syntax and logic paradigms.

\end{custommdframed}

\section{Discussion}
\subsection{Analysis of Repair Patterns}
\label{sec:patch_analysis}

In this section, we present a comprehensive analysis of the generated patches for successfully resolved issues, focusing on their complexity—measured by the average number of modified files and lines—relative to human-written patches.

Table~\ref{tab:patch_stats_discussion} compares the complexity of generated patches (Gen) by \ourmodel{} against human-written ground truth (GT). We observe distinct behavioral patterns across different benchmarks:
\begin{itemize}[leftmargin=*]
    \item \textbf{Compact Fixes in Complex Repositories:} On large-scale C++ repositories (Multi-SWE-bench), human patches are extensive (Avg. 70.16 lines changed across 2.24 files), often involving header restructuring or code cleanups. In contrast, \ourmodel{} generates highly compact, ``surgical'' fixes (Avg. 24.24 lines across 1.16 files). This suggests that in emergency response scenarios, the agent prioritizes \textit{minimal modification} to resolve the vulnerability, minimizing the risk of collateral regressions.

    \item \textbf{Verbose Fixes for Memory Safety Issues:} Conversely, in the domain of memory safety (SEC-Bench), \ourmodel{} tends to be more verbose than humans (NLA: 17.03 vs. 10.48). This stems from the \ourmodel{}, which encourages the injection of explicit guard clauses (\textit{e.g.}, redundant null checks) to block exploit paths. While less elegant than human refactoring, this strategy effectively enforces security invariants.
\end{itemize}

\begin{table*}[t]
    \centering
    \setlength{\abovecaptionskip}{0.1cm}
    \caption{Comparison of complexity between human-written (GT) and \ourmodel{} generated (Gen) patches.}
    \label{tab:patch_stats_discussion}
    \begin{tabular}{l c c c c c c c c c}
        \toprule
        \multirow{3}{*}{\textbf{Benchmark}} &
        \multirow{3}{*}{\textbf{\# Successful Cases}} &
        \multicolumn{2}{c}{\textbf{\# Changed Files}} &
        \multicolumn{6}{c}{\textbf{\# Changed Lines}} \\

        \cmidrule(lr){3-4} \cmidrule(lr){5-10}

         & & & &
        \multicolumn{2}{c}{\textbf{Added}} &
        \multicolumn{2}{c}{\textbf{Deleted}} &
        \multicolumn{2}{c}{\textbf{Total}} \\

        \cmidrule(lr){5-6} \cmidrule(lr){7-8} \cmidrule(lr){9-10}

         & & \multirow{-2}{*}{\textbf{GT}} & \multirow{-2}{*}{\textbf{Gen}} & \textbf{GT} & \textbf{Gen} & \textbf{GT} & \textbf{Gen} & \textbf{GT} & \textbf{Gen} \\
        \midrule

        SEC-Bench & 116 & 1.28 & 1.18 & 10.48 & 17.03 & 5.44 & 9.49 & 15.92 & 26.53 \\
        Multi-SWE (C++) & 37 & 2.24 & 1.16 & 33.00 & 16.76 & 37.16 & 7.49 & 70.16 & 24.24 \\
        PatchEval  & 131 & 1.01 & 1.45 & 13.99 & 13.14 & 4.95 & 3.57 & 18.94 & 16.71 \\
        \bottomrule
    \end{tabular}
    \vspace{-0.3cm}
\end{table*}

\subsection{Qualitative Analysis of Localization Mismatch}
\label{sec:qualitative_analysis}

In the analysis presented in Section \ref{sec:results}, we observed a phenomenon: a considerable prevalence of the $L_{\times}F_{\checkmark}$ category, where \ourmodel{} successfully repairs the vulnerability despite failing standard localization metrics (\textit{i.e.}, the modified files do not match the human-written ground truth). To further investigate this phenomenon, we conducted a manual analysis and identified two distinct repair patterns where the agent discovers \textbf{behaviorally equivalent} paths that differ spatially from the human-written ground truth:

\begin{itemize}[leftmargin=*]
    \item \textbf{Location Divergence:}  While human experts often modify a low-level utility function (\textit{callee}) to handle edge cases generically, \ourmodel{} tends to intercept the invalid input at the specific call site (\textit{caller}) or within a higher-level handler. Structurally different, this effectively blocks the vulnerability trigger path for the tested scenario.

    \item \textbf{Scope Divergence:} Human commits frequently encompass non-functional files such as header definitions. Since localization metrics require the predicted file set to be a superset of the ground truth, omitting these results in zero scores. \ourmodel{}, driven by the verification oracle, acts as a surgical agent, modifying only the \textit{core implementation files} required to mitigate the crash.
\end{itemize}

\noindent We illustrate these two patterns through specific real-world case studies below.

\noindent \textbf{Case 1: Functional Equivalence (CVE-2022-32414).}
An invalid object originates in the Promise handler (\texttt{src/njs\_promise.c}) but crashes only after propagating to the bytecode interpreter (\texttt{src/njs\_vmcode.c}). The human patch sanitizes at the \textbf{source}---the Promise handler---before the value enters the VM:
\begin{lstlisting}[language=C, title={\textbf{Ground Truth:} \textcolor{red}{src/njs\_promise.c}}]
// [Logic] Check validity BEFORE passing data to the iterator
 |\textcolor{diffgreen}{+}| if (!njs_is_valid(value)) {
 |\textcolor{diffgreen}{+}|     // Sanitize invalid input to undefined
 |\textcolor{diffgreen}{+}|     value = njs_value_arg(&njs_value_undefined);
 |\textcolor{diffgreen}{+}| }
// Pass the sanitized 'value' to the function call
ret = njs_function_call(vm, pargs->function, ..., value, ...);
\end{lstlisting}

\ourmodel{}, guided by the runtime stack trace, instead guards at the \textbf{destination}---the VM interpreter---intercepting the corrupted state before dereference:

\begin{lstlisting}[language=C, title={\textbf{\ourmodel{}:} \textcolor{red}{src/njs\_vmcode.c}}]
// [Logic] Switch source: Read from stable 'vm->retval' instead of unstable register
 |\textcolor{diffred}{-}| next = value2->data.u.next;
 |\textcolor{diffgreen}{+}| next = vm->retval.data.u.next;
// [Logic] Defensive Check: Abort if the iterator is corrupted
 |\textcolor{diffgreen}{+}| if (njs_slow_path(next == NULL || next->array == NULL)) {
 |\textcolor{diffgreen}{+}|     njs_internal_error(vm, "invalid iterator");
 |\textcolor{diffgreen}{+}|     goto error;
 |\textcolor{diffgreen}{+}| }
\end{lstlisting}

Although spatially distinct, the agent's patch creates a behaviorally equivalent safety invariant that effectively blocks the exploit path.

\noindent \textbf{Case 2: Scope Divergence (CVE-2019-13309, ImageMagick).} The human expert performed a multi-file repair, addressing memory leaks in both the primary command handler (\texttt{operation.c}) and the utility module (\texttt{mogrify.c}). \ourmodel{} correctly fixed the leak in \texttt{operation.c}---the path exercised by the PoC---but missed the corresponding leaks in \texttt{mogrify.c}, whose code paths were not triggered by the test oracle. This illustrates a limitation of test-driven repair: the agent resolves the active exploit path but may miss latent vulnerabilities in unexercised code.

\subsection{Threats to Validity}

\noindent\textbf{Internal Validity.} A primary concern is data leakage, including leakage during retrieval and leakage from the LLM’s pretraining data.
To mitigate the impact of the former as much as possible,
we enforced strict constraints during experience retrieval: we rigorously excluded any historical instances sharing the same unique CVE identifier and utilized timestamps to filter out any future code committed after the target vulnerability's disclosure.
As for the latter, although we cannot participate in model pretraining to effectively eliminate this factor, we avoid its impact on comparative results by using a consistent backbone LLM.
Spefically, in RQ1, we utilize the identical \texttt{DeepSeek-v3.2} backbone, thereby ensuring that performace improvements are attributable to our memory-guided architecture rather than model-inherent knowledge.

\noindent\textbf{External Validity.} This concerns the generalizability of our framework across languages and toolchains. A potential limitation is our reliance on the C++-specific \texttt{Iter\_grep} tool. However, our evaluation on PatchEval (RQ3) demonstrates that \ourmodel{} maintains state-of-the-art performance across Python, JavaScript, and Go even when \texttt{Iter\_grep} is replaced with standard text search, suggesting that effectiveness derives from the memory-guided mechanism rather than language-specific static analysis.
Regarding PoC dependence, \ourmodel{}'s refinement loop requires a vulnerability PoC for feedback. This is a standard AVR assumption rather than one we introduce: recent SoK studies~\cite{li2025sokeffectiveautomatedvulnerability} identify the PoC as the canonical input distinguishing AVR from general-purpose APR. All baselines in our evaluation share this requirement, and we grant OpenHands* the identical \texttt{check\_vul} tool to ensure equal PoC access.

\section{Related Work}
\subsection{Automated Vulnerability Repair}
The paradigm of Automated Vulnerability Repair (AVR) has evolved significantly to address the increasing complexity of software defects.
Early efforts primarily relied on \textit{template-based}~\cite{xuan2016nopol, huang2019using} or \textit{constraint-based}~\cite{liu2019tbar, zhang2022program, gao2021beyond, shen2025topology} techniques. While these methods prioritized semantic correctness, they suffered from the ``path explosion'' problem and struggled with complex logic flaws lacking formal specifications~\cite{zhang2020multiplex}.
Subsequent data-driven approaches, such as VRepair~\cite{fu2022vulrepair}, treated repair as a Neural Machine Translation (NMT) task but were often limited by dataset noise and shallow semantic understanding.

Recently, LLMs have reshaped the landscape, demonstrating exceptional proficiency in vulnerability detection, root cause analysis, and repair~\cite{kang2023explainable, ni2024next, fu2023chatgpt, liu2025securereviewerenhancinglargelanguage, thakur2024verrepair}.
To enhance LLM performance, literature has explored two primary directions.
One field focuses on augmenting context with static knowledge. Representative approaches like VulMaster~\cite{DBLP} inject expert knowledge (\textit{e.g.}, CWE~\cite{CWE} relationships) into the prompt to guide reasoning, while retrieval-based methods like CRepair~\cite{liu2024crepair} retrieve similar historical patches to serve as few-shot references. Others focus on abstracting patch patterns~\cite{cao2025enhancing} to improve generalization.
The other prioritizes dynamic validation. Earlier works utilized iterative feedback loops~\cite{xia2024automated} to filter invalid patches, while recent advancements like VulDebugger~\cite{liu2025agentdebugsdynamicstateguided} employ agents to actively inspect runtime program states, mimicking human debugging to pinpoint root causes.
However, existing methodologies often fail to effectively integrate retrieved knowledge with runtime feedback in repository-level environments, limiting their efficacy on resolving rigorous and complex repository-level vulnerabilities.

\subsection{Automated Issue Resolution}

Automated Program Repair (APR) has evolved significantly from traditional heuristics, such as encompassing search-based methods like GenProg~\cite{le2011genprog} and template strategies like TBar~\cite{liu2019tbar}, to advanced approaches driven by LLMs~\cite{xia2024chatRepair, yin2024thinkrepair, bouzenia2024repairagent, zhang2025ReinFix}.
Recently, the introduction of SWE-bench~\cite{jimenez2023swebench}, which evaluates systems on real-world GitHub issues, has catalyzed a paradigm shift toward repository-level resolution.
Driven by this benchmark, contemporary methodologies predominantly bifurcate into two architectures: interactive agents and fixed workflows. Interactive agents, such as representative OpenHands~\cite{wang2024openhands}, SWE-agent~\cite{yang2024sweAgent} and Aider~\cite{Aider}), employ LLMs as autonomous planning agents that perform end-to-end issue resolution through iterative interaction with development environments, tools, and codebases. These systems dynamically adapt their strategies based on execution feedback, enabling flexible problem-solving.  Fixed workflows (\textit{e.g.}, Agentless~\cite{xia2025agentless} and PatchPilot \cite{li2025patchpilot}) that rely on structured localization-repair pipelines with predefined stages. By decomposing the resolution process into explicit phase, such as issue understanding, fault localization, and patch generation, to resolve issues.

While effective for general functional bugs, these general-purpose resolvers often lack the domain-specific security priors required for vulnerability repair, failing to enforce critical defensive patterns—a limitation our framework addresses through memory-augmented reasoning.

\section{Conclusion}

In this paper, we proposed \ourmodel{}, a memory-guided agentic framework for repository-level vulnerability repair. By mimicking expert reasoning through a three-tier memory hierarchy and a feedback-driven refinement loop, our framework enables agents to recall historical patterns and learn from runtime trajectories. Extensive evaluation on SEC-Bench (C/C++), PatchEval (Python, Go, JavaScript), and Multi-SWE-bench (C++) demonstrate that \ourmodel{} achieves state-of-the-art vulnerability repair performance while effectively controlling computational cost, validating the efficacy of memory-guided reasoning for resolving complex, repository-level vulnerabilities across diverse programming languages. Looking ahead, we plan to explore more robust memory acquisition and consolidation strategies to further improve the robustness, scalability, and adaptability of agentic vulnerability repair systems.


\section*{Data Availability Statement}
Our code and data are available at \href{https://figshare.com/s/f2bbe7f7c8a759339368}{https://figshare.com/s/\\f2bbe7f7c8a759339368}.

\bibliographystyle{ACM-Reference-Format}
\bibliography{ref}

\end{document}